\documentclass[10pt,a4paper,twocolumn,prb]{revtex4}
\usepackage[]{graphicx}
\usepackage[]{pstricks} 

\begin{document}

\title{Study of Correlation Between Glucose Concentration and Reduced Scattering Coefficients in Turbid media using Optical Coherence Tomography}

\author{R. Poddar}
\altaffiliation[Also at ]{
Department of Biotechnology, Birla Institute of Technology, Mesra, Ranchi,
835~215 India}
\affiliation{Applied Optics Laboratory, Department of Applied Physics, Birla Institute of Technology, Mesra, Ranchi, 835~215 India}
\author{S. R. Sharma}
\affiliation{Applied Optics Laboratory, Department of Applied Physics, Birla Institute of Technology, Mesra, Ranchi, 835~215 India}
\author{P. Sen}
\affiliation{Department of Physics, Devi Ahilya University, Indore  452~001 India}
\author{J. T. Andrews}
\altaffiliation{Permanent Address Department of Applied Physics, Shri
G S Institute of Technology \& Science, Indore, 452 003
India} 
\email{ Email: jtandrews@sgsits.ac.in}
\affiliation{Applied Optics Laboratory, Department of Applied Physics, Birla Institute of Technology, Mesra, Ranchi, 835~215 India}

\begin{abstract}
Noninvasive, non-contact and \textit{in vivo} monitoring blood glucose is a long needed pathology tool for saving patients from 
recurring pain and hassle that can accompany conventional blood glucose testing methods. Optical coherence tomography known for its high axial resolution imaging modality is adopted in this article for monitoring glucose levels in tissue like media non-invasively. Making use of changes in reduced scattering coefficient due to the refractive-index mismatch between the extracellular fluid and the cellular membranes and armed with a theoretical model, we establish a correlation between the glucose concentration and reduced scattering coefficient. The scattering coefficients are extracted from the deconvoluted interference signal by using Monte-Carlo simulation with valid approximations. A program code using NI LabVIEW$^{TM}$ is developed for automation of the experiment, data acquisition and analysis.
\end{abstract}

\maketitle
\section{INTRODUCTION}
The method of optical coherence tomography (OCT) was first introduced by Huang and colleagues. It has been widely applied to medical imaging and diagnostics \cite{Huang}. OCT have the capability to acquire two- and three-dimensional tomographic images in biological tissues. These applications are limited by penetration depth, cross sectional area, dynamic range and signal to noise ratio (SNR). However, it is successfully applied to transparent ocular organs where light scattering is minimum. Multiple scattering, which becomes dominant at large depths, is the fundamental limitation preventing OCT from reaching a large probing depth in turbid media \cite{Fujimoto}. OCT is widely used as a biomedical imaging modality. We extent this idea towards a new direction where tomo-
graphic (cross sectional) imaging is not the prime goal,
instead we measure the optical properties of stratified
media with better accuracy.

The optical properties themselves can potentially provide information to monitor tissue metabolic status or to diagnose disease. Optical approaches to study turbid media in the presence of chiral \cite{Wang} components have generated interest because of their potential use in noninvasive glucose monitoring for diabetes patients. Light scattering occurs in tissues because of the mismatch of index of refraction between the extra cellular fluid (ECF) and the membranes of the cells composing the tissue. In the near-infrared region, the index of refraction of the ECF is $n_{ECF} \approx$ 1.348-1.352 while the index of refraction of the cellular membranes and protein aggregates is in the range $n_{cell} \approx$ 1.350-1.460 \cite{data1,data2,data3}. It is well known that adding sugar to water increases the index of refraction of the solution. Similarly, adding glucose to blood in turn raise the refractive index of the ECF, which will cause a change in the scattering characteristic of the tissue as a whole. Hence, tissue glucose levels are correlated with scattering coefficients based on changes in the refractive index of extra cellular fluid. Of late, measurements on light scattering by blood shows promising correlation between blood glucose and reduced scattering coefficient \cite{Mai}. On the other hand, the light scattering is not influenced by the red blood cells and other chemical composition of blood. At the same time, monitoring of glycemic status in patients with diabetes requires determination of blood glucose concentration. Significant efforts have been made by several groups in the past few decades to develop a bio-sensor for noninvasive blood glucose analysis.

Different optical approaches were proposed to achieve this goal. These approaches include polarimetry, Raman spectroscopy, near-infrared (NIR) absorption and scattering, and photo-acoustics \cite{Cote,Goetz,Pan1,Gabriely,MacKenzie}. Although these techniques are promising, they have limitations associated with low sensitivity, accuracy and insufficient specificity of glucose measurements at physiologically relevant levels.
Of late, Larin et al \cite{Esen, Larin} have proposed a possible change in  slope of OCT signal due to changes in glucose in blood by optical coherence tomography. However signal analysis procedure is quite complex and less accurate. They used a used a linear fit model to deduce OCT slope  by using the least-squares method. We employ a three
step procedure which is entirely different from their technique: (i) OCT signal is deconvoluted from the source
function, (ii) extraction of optical properties of the turbid media using Monte Carlo simulations with few valid
approximations and (iii) interpret the changes in glucose
concentration from the measured optical parameters.

Some theoretical models were developed to understand the governing physical process and to better interpret the OCT signal in highly scattering media. Pan et al, established the relationship between the path-length resolved reflectance signal and the OCT signal using linear system theory \cite{Pan1,Pan2}. Monte Carlo technique was employed to simulate the path-length resolved reflectance but could not be able to separate the effects of the singly scattered light and the multiply scattered light.\cite{Pan} The OCT signal was split into (i) summation of singly back-scattered light (coherent) and (ii) multiply scattered light (partially coherent). The effect of multiple scattering on the formation of speckle patterns and the degradation of image contrast were demonstrated. In reality, light scattering in turbid media is a complex process, and it is only an approximation to assume that the OCT signal is from single back-scattering alone. Photons still contribute to the OCT signal after a limited number of scattering events. The multiple scattering effects are clearly demonstrated in terms of the spreading of the point spread function (PSF).

In the present paper, we employ Optical Coherence Tomography to monitor the reduced scattering coefficient for different values of glucose concentration in an aqueous solution with Intralipid as the scatterers. Monte-Carlo simulation technique with valid approximations is adopted to understand the contribution of the multiple-scattered light obtainable from OCT. The experimental observation are supported by the theoretical analysis based upon the transport theory. A strong correlation between reduced scattering coefficient and glucose concentration is established. Maximum measurement error
of 5\% is observed at hypo-glycemic range.
\section{THEORY}

Michelson Interferometer is the basis of any OCT setup (as schematically shown in Fig. 1), a motor controlled reference mirror and sample with focusing assembly are kept in two arms of the interferometer. Light scattered from the sample arm and the light reflected from the reference arm interfere and is detected by a photodiode. The Doppler frequency  generated by constant scanning speed of the mirror is modulated by a coherence function of the low coherent light source. A bandpass filter centered at the Doppler frequency acts as a coherence gate for signal detection. 
       \begin{figure}[h]
        \begin{center}
        \includegraphics[height=6cm]{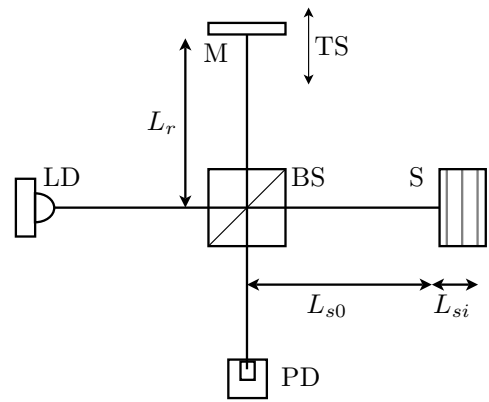}
        \end{center}
        \caption{\sf Schematic of the experimental setup. SLD - Superluminiscent                    diode, M - Mirror, BS - Beam Splitter cube, TS - stepper                        motor controlled translation stage, PD - photodiode,
                       S~-~sample}                                      
        \end{figure}
The sample in a OCT setup used to be a turbid medium such as tissue, plants, composites, etc. Light scattered from a turbid medium may be broadly classified under two categories, viz., (i)~the least scattered light which undergoes only single or very little scattering and (ii)~diffusely scattered light, which undergoes multiple scattering events. From the principle of classical theory of scattering, least scattered light maintains coherence whereas multiply scattered light lose coherence. Due to finite width of the coherence gating employed in OCT, multiple scattered light with path length difference falling within the coherence length of the source are detected. The optical irradiance at the detector is superimposition of all the light fields reflected from within the scattering sample and the reference mirror and is given by,
\begin{equation}
I_{d}(\tau) = \left\langle\left| \int^{\infty}_{Ls} E'_{s}(t,L_{s})dL_{s} + E_{r}(t,\tau)\right|^2
 \right\rangle, \label{Idet}
\end{equation} 
where, $\tau \;[\;= (L_s-L_r)/c = \Delta L /c\;]$ is the time delay corresponding to the round trip optical path length between two beams, $L_{s} [ = L_{s0}+\sum_{i=0}^{\infty}{L_{si}}]$ takes care of round trip path length to the sample surface and the total path length within the sample that accumulates during each scattering. $E'_{s}(t,L_{s})$ is path length resolved field intensity.
The first part of eq. (\ref{Idet}) is important for the present study since it contains information regarding optical properties of turbid media. 

OCT signal obtainable from a turbid media is a convolution of path-length resolved diffuse reflectance (arises from turbidity of the media) and the low-coherence function arises from coherence property of laser source. Accordingly, OCT signal may be rewritten as,
\begin{equation}
I_{d} (L_{r}) = 2\sqrt{I_{s}I_{r}}[R(L_{s})^{1/2} \otimes C(L_{s})].\label{Id}
\end{equation}
Where, $I_{d}(L_{r})$ is the OCT signal detected by a photodiode, $I_{s}$ and $I_{r}$ are the signals from sample and reference arms respectively, $R(L_{s})$ is path-length resolved diffuse reflectance and $C(L_{s})$ is low-coherence function of the source.
In order to extract the scattered light ($R(L_s)$) from $I_d$, it is deconvoluted with the coherence function $C(L_s)$. The scattering coefficients are obtained by fitting the deconvoluted signal with the following equation obtainable from transport theory
\begin{equation}
\nabla^2 U_d(r) - K_d^2 U_d(r) = -\frac{3}{4\pi}\rho \mu_{tr}P_0\delta(r). \label{difeq}
\end{equation}
Here, $U_d(r)$ is the average diffuse intensity, $P_0$ is the total radiating power and $K_d \; (= \sqrt{3\rho^2 \mu_a \mu_{tr}})$ is a fluid constant which varies with concentration of the fluid. Here, $\mu_{a},\,\mu_{s}$ and $\mu_{tr}$ are the absorption, scattering and transport coefficients, respectively and $\rho$ is the number density of scatterers. 

Light scattering in turbid media depends strongly on the value of anisotropy parameter $g$, which is the average cosine of the scattering angle. Total forward scattering occurs for $g$ = 1 while isotropic scattering occurs for $g$ = 0. In tissue like media the light scattering is generally highly forward peaked since the anisotropy parameter $g \geq $  0.8.  Since OCT signal composed of backscattered photons, the scattering coefficient is replaced with the reduced scattering coefficient \[\mu_{s}^{\prime }= (1-g) \mu_{s}\] instead of the conventional scattering coefficient, to take care of changes arising due to the anisotropy in the system. Also, the wavelength of light source can be appropriately chosen to minimize the light absorption such that $\mu_{a}^{2}<< \mu_{s}^{2}$.  \cite{Ishi} Accordingly, the transport coefficient is now redefined in terms of the absorption and scattering coefficients as%
\begin{equation}
\mu_{tr}=\mu_{a}+(1-g)\mu_{s}=\mu_{a}+\mu_{s}^{\prime }.
        \label{mutr}
\end{equation}

After simplification, the solution of Eq. (\ref{difeq}) is found to be
\begin{equation}
\frac{r}{r_0}\frac{U_d(r)}{U_{d0}(r_0)} =\exp[-K_d(r-r_0)]. \label{sol}
\end{equation}
From the above equation and radial distribution of diffuse light relative change of scattering coefficient and glucose concentration can be measured.
\textbf{}  \section{Experiment}

We use an IR superluminescent diode (SLD) emitting at 810 nm (Hamamatsu) as a light source. The wavelength of the light source is appropriately chosen such that the tissue absorption can be ignored \cite{Maier}. The scattered light is detected by a photodiode (Hamamatsu) and the signal is measured using a lock-in amplifier (EG$\&$G 7265).

The motor controlled stages, data acquisition system
and the lock-in amplifier are controlled through PC. 
After deconvolution of the OCT signal ($I_d$) with reference to the source function ($C(L_s)$) of light source, diffuse reflectance signal ($R(L_s)$) is obtained. It is further used with transport theory to calculate the reduced scattering coefficient of the turbid media. The data collected for different concentration of glucose are analyzed and the scattering coefficient is estimated for each measurement {with valid Monte-Carlo approximations \cite{Poddar}.
                                                                                                \begin{figure}[h]
                        \centering{\includegraphics[height=6cm]{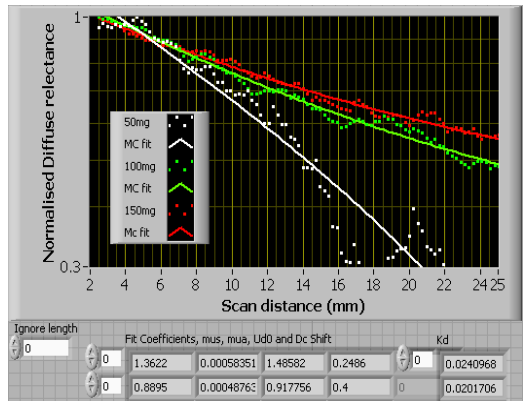}}
                        \label{labview}
                       \caption{\sf Front-panel of Monte-Carlo simulation program written using LabVIEW. The dotted curves are obtained from experimental data while the solid lines are fit data  using simulation. The simulation constants are displayed at the bottom. The results are displayed after 1000 iteration (optimized).}                                                                         \end{figure}

\textbf{Sample description:} Glucose solutions prepared with different concentration were mixed with turbid media (Intralipid™ 0.1 v/V, values of scattering coefficient and anisotropy parameter that are used for the tissue phantom are 50cm$^{-1}$ and 0.8 respectively), used as sample. The exact values of the glucose concentration used in experiment ranges from 20mg/dl to 2000mg/dl to establish the correlation between the hypoglycemic stage ($\leq$ 64 mg/dl) and hyperglycemia stage ($\geq$ 100 mg/dl). One should note here that the normal blood glucose level range is 64mg/dl to 100mg/dl. The time allotted for settlement and interaction of glucose with Intralipid is 2 minutes. Since, the reduced scattering coefficient of glucose are not altered immediately after the addition glucose. 
\section{RESULTS and DISCUSSIONS}
In order to carry out the simulation a program code is written using LabVIEW$^{TM}$ with 1000 iterations (with option for $N$ number of iterations). The program window is shown in Fig. 2. The program makes use of eqs. (\ref{Id}) - (\ref{sol}) for simulation and data fitting. The method adopted by Larin et al \cite{Esen, Larin} to monitor glucose concentration in
tissue phantom, obtains tomographic images (2D), which
are averaged to 1D distribution of light in depth. The
1D distributions were plotted in a logarithmic scale to
find the slope of the distribution at different depths using
least-squares method. Which leads a less accurate results
with large error in measurement. We adopted a different
direction, where the OCT signal is deconvoluted from
the source function of the SLD. The deconvoluted signal
is fitted using MC simulation with the PSF as defined
earlier. The following parameters are estimated from the
program code: $\mu'_s$, $\mu_a$ and fit constants arising due to experimental dark noise and background noise. 

The OCT signal obtained for different concentration
of glucose are exhibited in Fig. \ref{oct} for a fixed concentra-
tion of intralipid. With increase in glucose concentration the amplitude of the OCT signal decreases.\cite{Pan} Also, one can notice that the Gaussian width of the curve increases. However, one can not extract much information about the scattered light from these curves, since a strong coherence function from the low coherent light source is convoluted on the scattered signal. The coherence data obtained from an OCT signal of the non-scattering medium like water is used 
         \begin{figure}[ht]
         \centering
         \includegraphics[height=6cm]{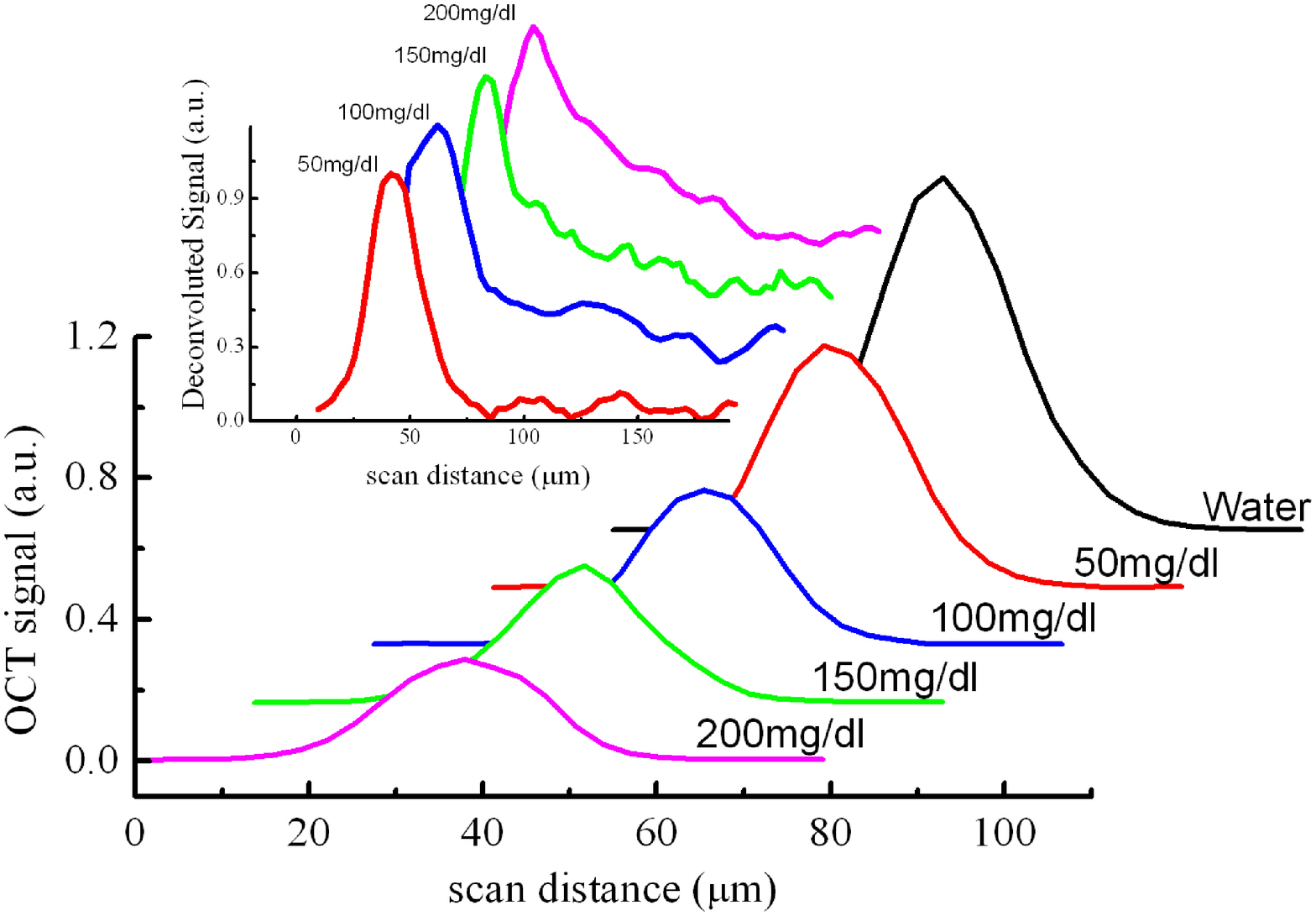}
      \caption{\sf OCT signal for different glucose concentrations. For larger concentrations, the amplitude of signal decreases, while the width of the curve increase. The deconvoluted signals are shown in inset. The OCT signal obtained for water is taken as the source coherence function. The deconvoluted signals are normalised. } \label{oct}\end{figure}
as the source function ($C(L_s)$). Other OCT signals obtained for different concentrations of glucose are deconvoluted and shown in the inset of Fig. \ref{oct}. The deconvoluted signals contain more information about the light scattering. With increasing concentration the area as well as the tail part of the signal increases. This is a clear indication of increasing value of reduced scattering coefficient with decreasing glucose concentration.  
                                       \begin{figure}[ht]
                                       \centering
                                       \includegraphics[height=5.5cm]{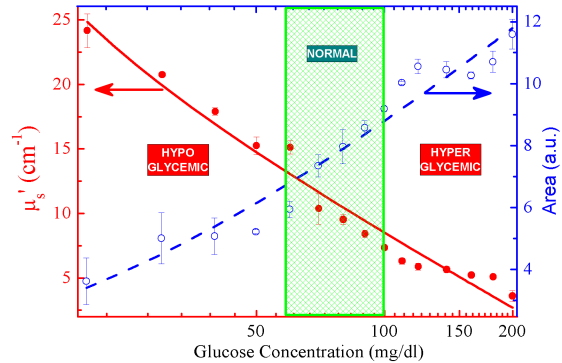}
                                       \caption{\sf Semilog plot of reduced scattering coefficient and curve area with glucose concentration. These data extracted from Monte-Carlo simulation and fitting it to the experimental data. The error bars are obtained after 25 measurements. } \label{glu}
                                       \end{figure}
The measured scattering coefficients obtained after Monte-Carlo simulation are depicted as a semilog plot in Fig. \ref{glu}. The curve (solid circles) exhibiting the nature of reduced scattering coefficient for different concentrations of glucose demonstrates a near logarithmic nature. In hypo-glycemic region it shows a sharp change while in the hyper-glycemic region it has smaller slope leading to less accuracy. A similar nature is also observed while measuring the area of the deconvoluted signal (open circles). This also experiences a near logarithmic behavior. Repeated measurement shows that the slope of the curves is a constant. Also, addition of more or less intralipid does not change this behavior. However, in order to find the value of unknown concentration of glucose, one needs an initial value at particular concentration of glucose. Since, the slope remains a constant, this method could be better solution for non-invasive, non-contact, in-vivo monitoring of blood glucose concentration. In order to realize this technique as a clinical tool more efforts are required.

\section{CONCLUSION}
To conclude, armed with light scattering technique and optical coherence tomography, we make an attempt to study a correlation between, glucose concentration and reduced light scattering coefficient which is better and accurate with
than Larin et al \cite{Esen, Larin}. A semilog plot of glucose concentration with reduced scattering coefficient suggests a linear relation. The value of reduced scattering coefficients are determined using Monte-Carlo Simulation. The technique promises to be a possible clinical tool for non-invasive and non-contact measurement of blood glucose. However, to predict exact value of blood
glucose without feeding any initial parameter, the equip-
ment needs large number of training in a pathology lab-
oratory.

\begin{acknowledgments}The authors thank DST, DRDO (LSRB) and AICTE for financial Support. SRS thank CSIR, New Delhi for fellowship.
\end{acknowledgments}

\end{document}